\documentclass[a4paper, 10pt, conference]{IEEEtran}

\usepackage{booktabs} % For formal tables
\usepackage{todonotes}
\usepackage{dirtytalk}
\usepackage{url}
\usepackage{graphicx}
\usepackage{subcaption}
\usepackage{balance}

\newcommand{\totalcount}{102'028}
\newcommand{\pscount}{11'142}
\newcommand{\derivatecount}{91'886}
\newcommand{\totalsize}{40.2} %40.166
\newcommand{\avgderivates}{7.9} %7.93

\begin{document}
\title{The PS-Battles Dataset --- an Image Collection for \\ Image Manipulation Detection}

\author{Silvan Heller, Luca Rossetto and Heiko Schuldt \\
Department of Mathematics and Computer Science\\
University of Basel, Switzerland\\
\{firstname.lastname\}@unibas.ch
}

%\keywords{Image Manipulation, Near Duplicate Detection, Content-based Retrieval}

\maketitle

\begin{abstract}
The boost of available digital media has led to a significant increase in derivative work. 
With tools for manipulating objects becoming more and more mature, it can be very difficult to determine whether one piece of media was derived from another one or tampered with. As derivations can be done with malicious intent, there is an urgent need for reliable and easily usable tampering detection methods. However, even media considered semantically untampered by humans might have already undergone compression steps or light post-processing, making automated detection of tampering susceptible to false positives.
In this paper, we present the PS-Battles dataset which is gathered from a large community of image manipulation enthusiasts and provides a basis for media derivation and manipulation detection in the visual domain. The dataset consists of \totalcount{} images grouped into \pscount{} subsets, each containing the original image as well as a varying number of manipulated derivatives. 
\end{abstract}

\section{Introduction}
\label{sec:introduction} 
Media creation in the digital age is an increasingly distributed process where the line between producer and consumer of media gets more and more blurry. In such a prosumer~\cite{TofflerWave} ecosystem, material is often sourced from various places, reused, manipulated, and shared several times which makes proper source attribution a difficult task.
As image manipulation software evolves and its use becomes more widespread, there is a need to verify the effectiveness of manipulation detection algorithms against images created by a diverse spectrum of tools, manipulations and proficiency in manipulation. However, not all tampering changes the semantic content of the image. Detecting JPEG-Compression and post-processing are not necessarily as relevant to users as manipulations which change content or context of an image. While a lot of work has been done to detect the presence of manipulations, we are not aware of out-of-the-box classifiers for tampered images.
One promising avenue are machine learning techniques which however require large amounts of data to work. We hope that by providing a large extendable dataset, research on automated classifiers can be stimulated.

\begin{figure*}
\centering
	\begin{subfigure}[t]{0.3\textwidth}
        \includegraphics[width=\textwidth]{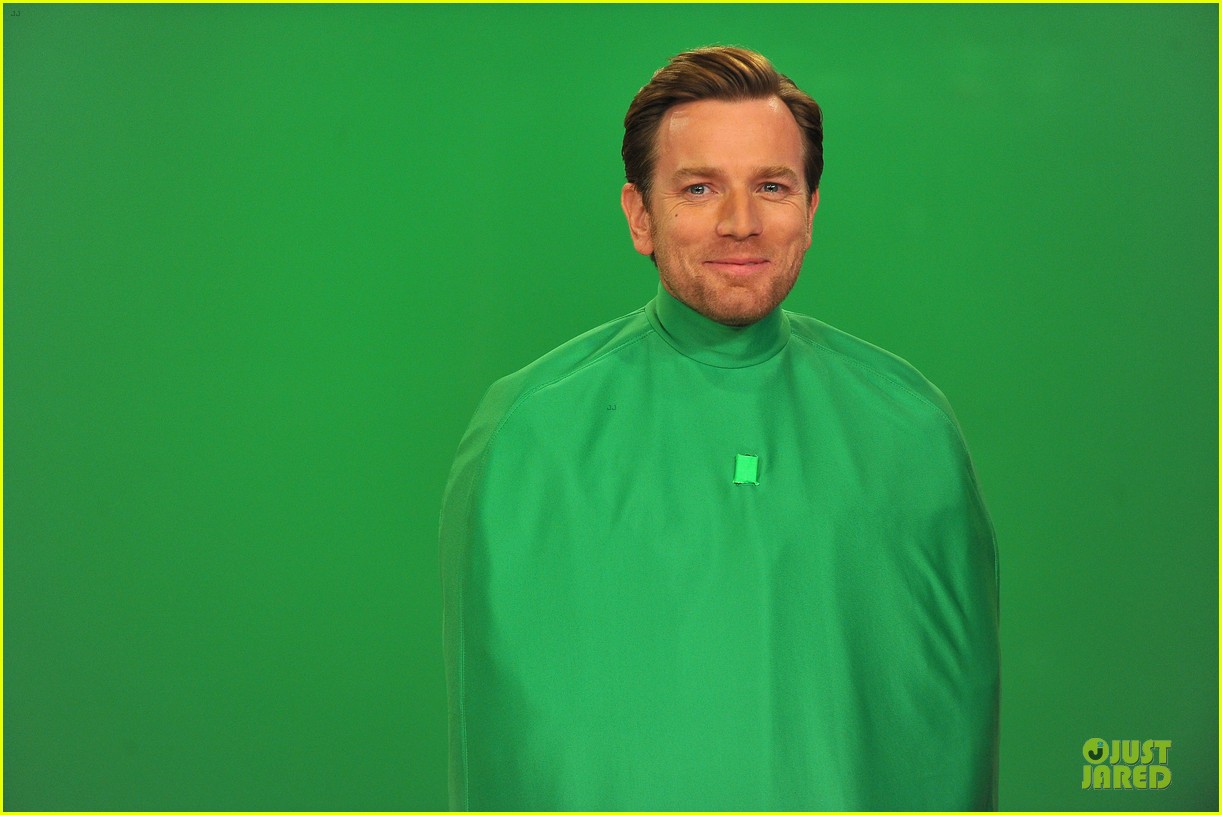}
        \caption{'Original' image: \url{http://i.imgur.com/sojfXm7.jpg}}
        \label{fig:orig1}
    \end{subfigure}   
    \begin{subfigure}[t]{0.3\textwidth}
        \includegraphics[width=\textwidth]{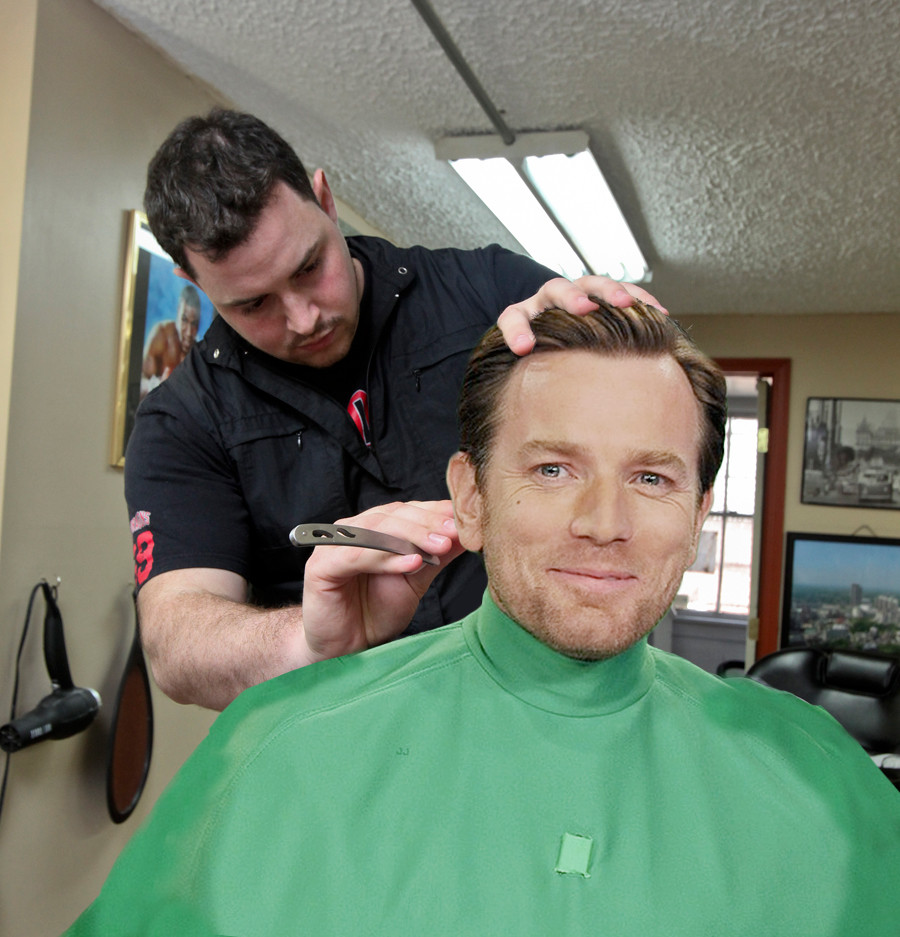}
        \caption{'at the barber' by 'totalitarian\_jesus': \url{https://imgur.com/tSwrvVn}}
        \label{fig:ps11}
    \end{subfigure}
    \begin{subfigure}[t]{0.3\textwidth}
        \includegraphics[width=\textwidth]{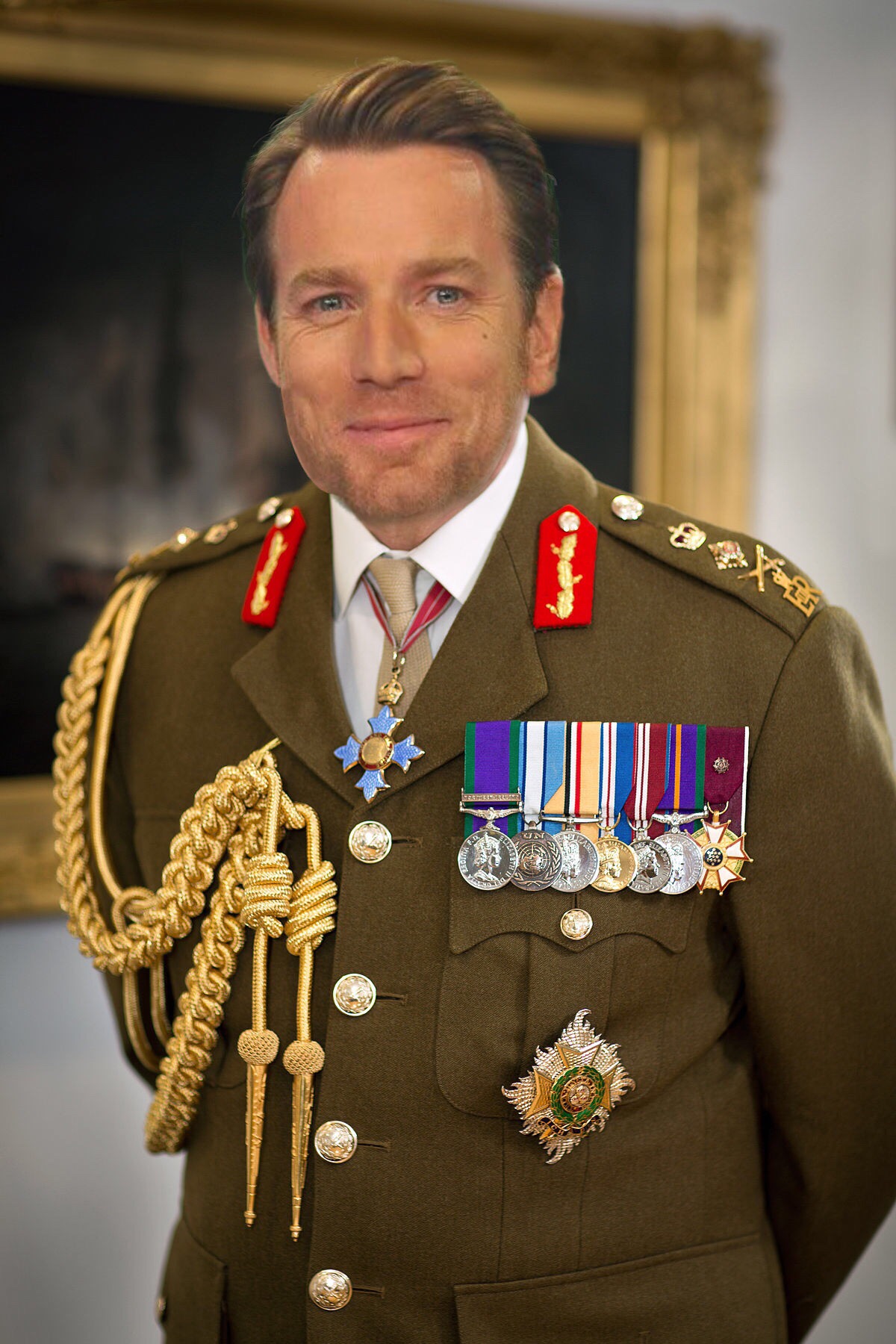}
        \caption{'General Kenobi!' by 'mandal0re': \url{https://imgur.com/adTHYDh}}
        \label{fig:ps12}
    \end{subfigure}
    \begin{subfigure}[b]{0.3\textwidth}
        \includegraphics[width=\textwidth]{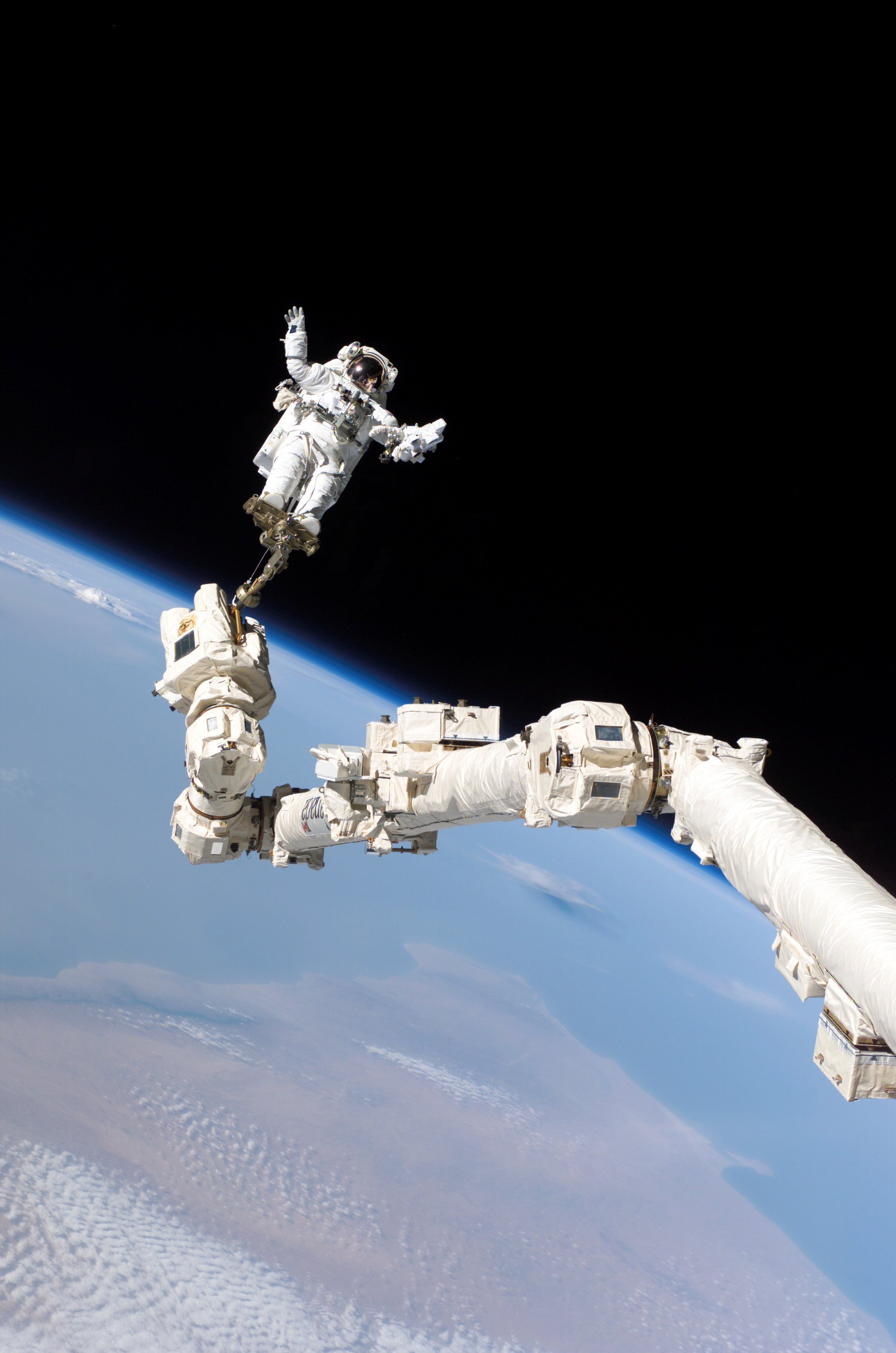}
        \caption{'Original' image from \url{https://i.imgur.com/0ZlLZUL.jpg}}
        \label{fig:orig2}
    \end{subfigure}
    \begin{subfigure}[b]{0.3\textwidth}
        \includegraphics[width=\textwidth]{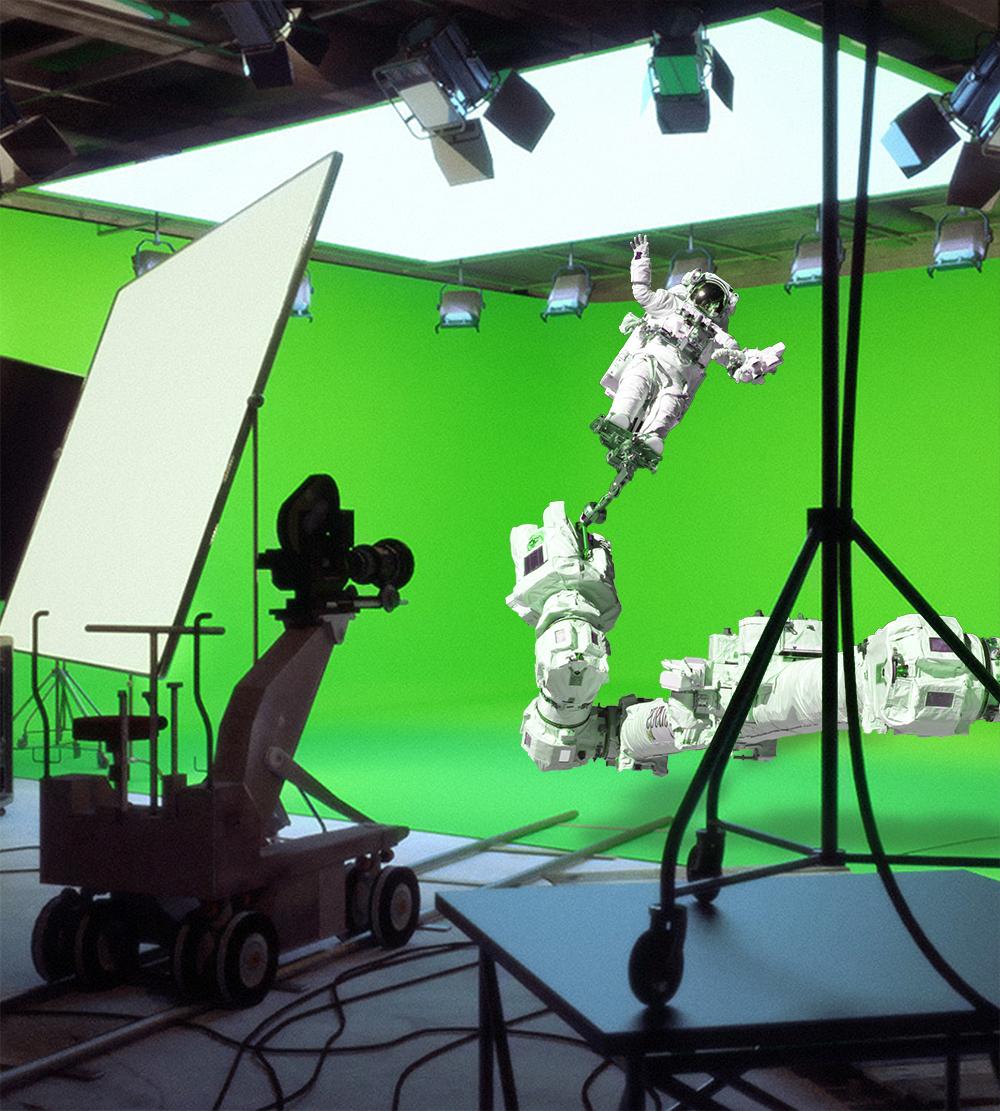}
        \caption{'Lies I tell you' by 'GalacticBystander': \url{https://imgur.com/zmEVgSR}}
        \label{fig:ps21}
    \end{subfigure}
    \begin{subfigure}[b]{0.3\textwidth}
        \includegraphics[width=\textwidth]{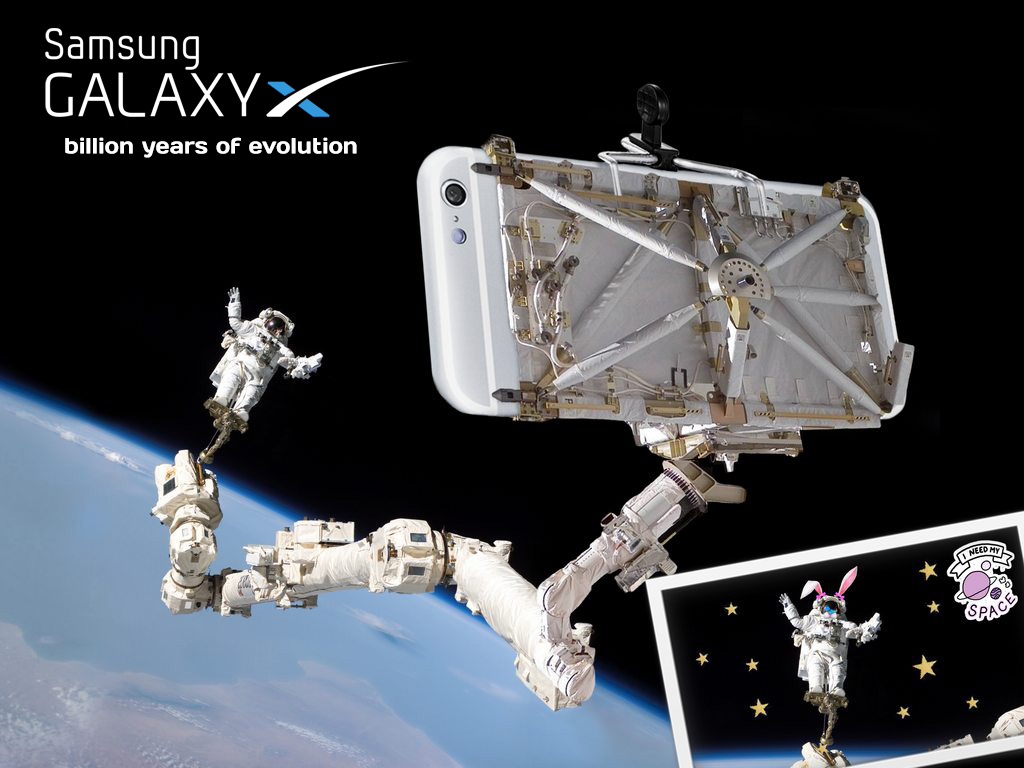}
        \caption{'Cosmic Selfie' by '-JenM-': \url{https://imgur.com/eWlT4WU}}
        \label{fig:ps22}
    \end{subfigure}

\caption{Examples of Original and Derivative Images}\label{fig:examples}
\end{figure*}

To further research in the area of modification detection, derivation detection as well as source identification in the visual domain, we present the \emph{PS-Battles} dataset. It is comprised of images sourced from the popular \emph{photoshopbattles} subreddit\footnote{\url{https://www.reddit.com/r/photoshopbattles/}} which is home to a large community of both amateur and professional digital artists who regularly hold contests in digital image manipulation. For every submitted original image, the community creates several, often humorous derivative images or so-called \emph{photoshops} which are then judged by other members of the community. Examples of such original images and the community-created derivatives can be seen in Figure~\ref{fig:examples}. Reddit\footnote{\url{reddit.com}} is one of the most popular websites in the world, as of early 2018 ranking 7$^{th}$ globally and 4$^{th}$ in the US\footnote{\url{https://www.alexa.com/siteinfo/reddit.com}}. The \emph{photoshopbattles} community has 12.9~million subscribers, which makes it the 33$^{rd}$ largest community on reddit\footnote{\url{http://redditmetrics.com/r/photoshopbattles}}.

The presented dataset contains \pscount{} subsets consisting of the original image as well as several corresponding photoshops for a total of \totalcount{} images.
For every derivative image, the dataset contains additional metadata about the image's author, the time of its creation, and its reception within the community.
Since the photoshopbattles community is quite active, the dataset is extensible over time. 

The remainder of the paper is organized as follows: Section~\ref{sec:related} reviews related work. Section~\ref{sec:dataset} introduces the PS-Battles dataset and details some of its properties, and Section~\ref{sec:conclusion} concludes.

\section{Related Work}
\label{sec:related}
Both visual near duplicate detection as well as image tampering detection have gained in interest over the past years. The authors of~\cite{birajdar2013digital} provide an overview of image tampering detection techniques with a focus on passive or blind image forgery detection methods. They observe a \say{lack of established benchmarks and of public testing databases which evaluates the actual accuracy of digital image forgery methods.}~\cite{birajdar2013digital}. Most research reviewed either evaluates against automatically generated forgeries or a small set of manually created derivates. The authors are not aware of any large dataset for image tampering detection.
\cite{christlein2012evaluation}, for instance, focuses on Copy-Move Forgery Detection and evaluates against a dataset with 48 base images and derivatives.
\cite{rocha2011vision} also provides an extensive overview of the field of digital forensics. The authors also note that \say{the evaluation of existing and new algorithms must be improved. The analysis of detection results in nearly all papers surveyed lacks the rigor [...], making the assessment of their utility difficult.}~\cite{rocha2011vision}. Both \cite{birajdar2013digital} and \cite{rocha2011vision} also provide an extensive overview of single / double JPEG compression detection. For originals in our dataset, there is no guarantee that no JPEG compression artifacts will be found. We feel this is an advantage as it better represents the real-world use case tampering algorithms will face.
The CASIA\footnote{\url{http://forensics.idealtest.org/}} dataset~\cite{dong2013casia} is probably the most similar to our proposed dataset and contains 12'614 images, of which 5'123 are tampered with. The advantage of our approach is however that the community we source our content from is still active, so it is very likely that the dataset further grows in the future. Additionally, the untampered images of the CASIA dataset are completely untouched which is an unrealistic assumption for real-world applications.
There are other datasets, for instance the Copydays dataset~\cite{jegoucopydays,jegou2008hamming}, which is generated using automated artificial attacks and used for copy detection.
The CISDE~\cite{ng2009columbia} dataset provides 1'845 spliced picture blocks with a fixed size of 128 $\times$ 128. The spliced blocks lack context however, making them semantically meaningless.

Recently, the RAISE dataset~\cite{dang2015raise} was introduced which contains 8'156 untampered high-resolution raw images. There, the authors also discuss the lack of a comprehensive large-scale dataset. In a recent paper focusing on image provenance~\cite{moreira2018image}, also a dataset related to the photoshopbattles community is introduced. Their dataset however contains only 10'421 images and is focused comment chains, where derivations of derivations are made.

The fact that our dataset consists of image derivatives which have been generated using current industry-standard image manipulation techniques can in certain instances also be considered a disadvantage. \cite{huang2017densely} for example has shown that it is possible to generate modifications of faces using deep learning which are visually practically indistinguishable from original images. Such modifications have however not yet reached the main stream and have not made their way in any image tampering dataset we are aware of.

\section{Dataset Description}
\label{sec:dataset}

\begin{figure}[t]
\centering
\includegraphics[width=0.5\textwidth]{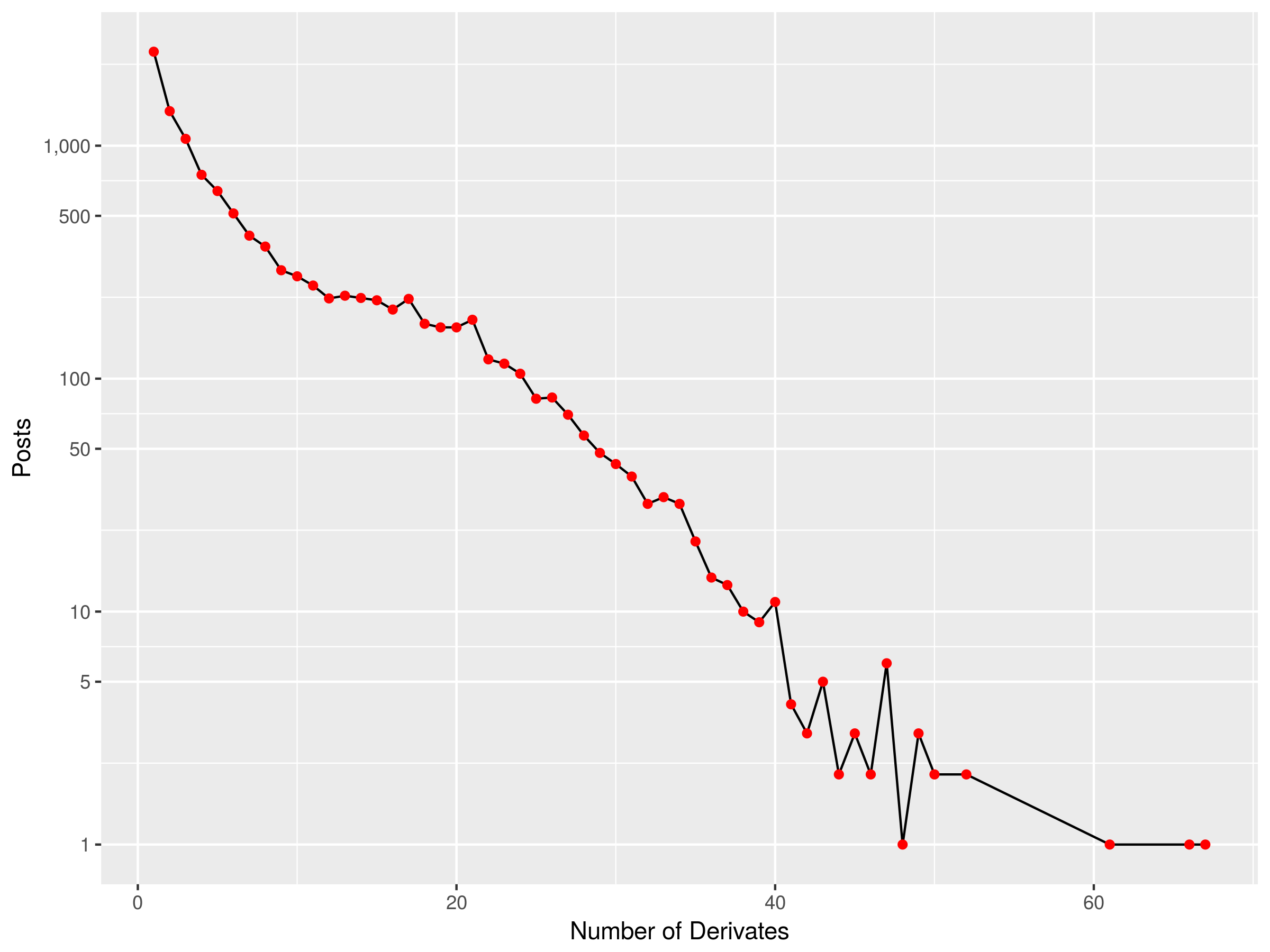}
\caption{Derivates per Original}
\label{fig:derivation}
\end{figure}

The following section describes the structure and properties of the proposed PS-Battles dataset and also presents in detail how it was collected. The dataset itself can be obtained from GitHub via \url{https://github.com/dbisUnibas/ps-battles}.

\subsection{Overview}
On average, the dataset contains \avgderivates{} photoshops for every original image.
Figure \ref{fig:derivation} shows the distribution of the number of derivate images per original.

\begin{figure}[t]
\centering
\includegraphics[width=0.5\textwidth]{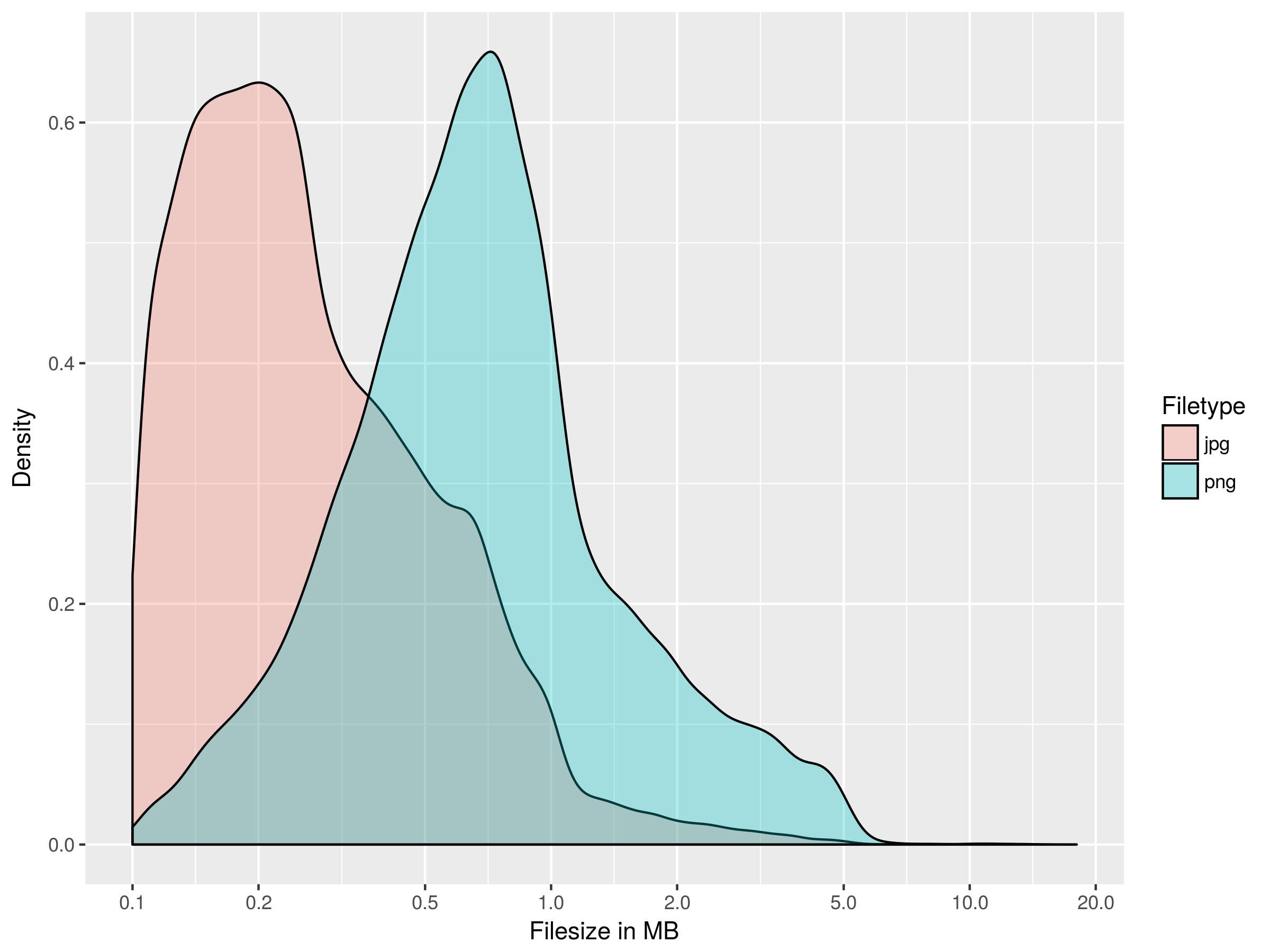}
\caption{Size per File Type}
\label{fig:filesize}
\end{figure}

As we can see on Figure \ref{fig:derivation}, there are a lot of posts with a small number of derivates and even the most popular posts do not exceed 67 derivates. This is expected as derivate creation is heavily moderated.
The combined size of all images in the dataset is \totalsize{}~GB. The distribution of the sizes of the individual files by file type can be seen in Figure~\ref{fig:filesize}.

\begin{figure}[t]
\centering
\includegraphics[width=0.5\textwidth]{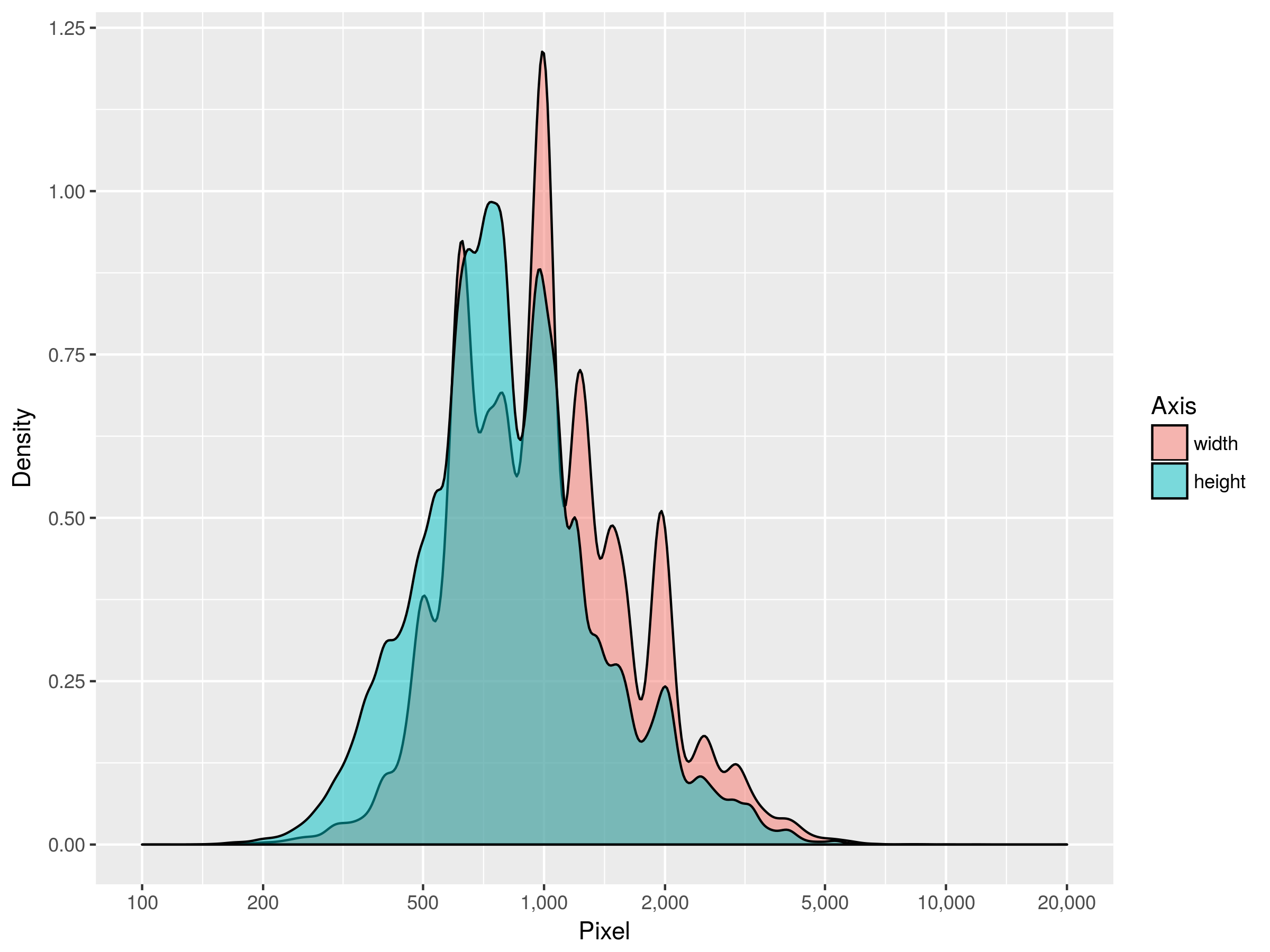}
\caption{Image Width and Height Density}
\label{fig:imgsize}
\end{figure}

As the dataset is community-generated, images vary in resolution and aspect ratio. Figure \ref{fig:imgsize} shows the distribution of width and height of the images in pixels.

Throughout the dataset, image height spans the range from 136 pixels for the smallest to 20'000 pixels  while image width goes from 68 pixels in the most narrow image to 12'024 pixels in the widest.
We feel this diversity in image dimensionality is beneficial as it makes the dataset more challenging.

\subsection{Collection Method}

In order to compile the dataset, we used a publicly available dump of reddit content\footnote{\url{http://files.pushshift.io/reddit/}} from which all posts and corresponding comments of the photoshopbattles subreddit were extracted. The moderators of the subreddit ensure that every post contains a link to the original image and every top-level comment contains a link to a photoshop. Lower level comments are then used by the community to discuss the manipulated images. 

\begin{figure}[t]
\centering
\includegraphics[width=0.5\textwidth]{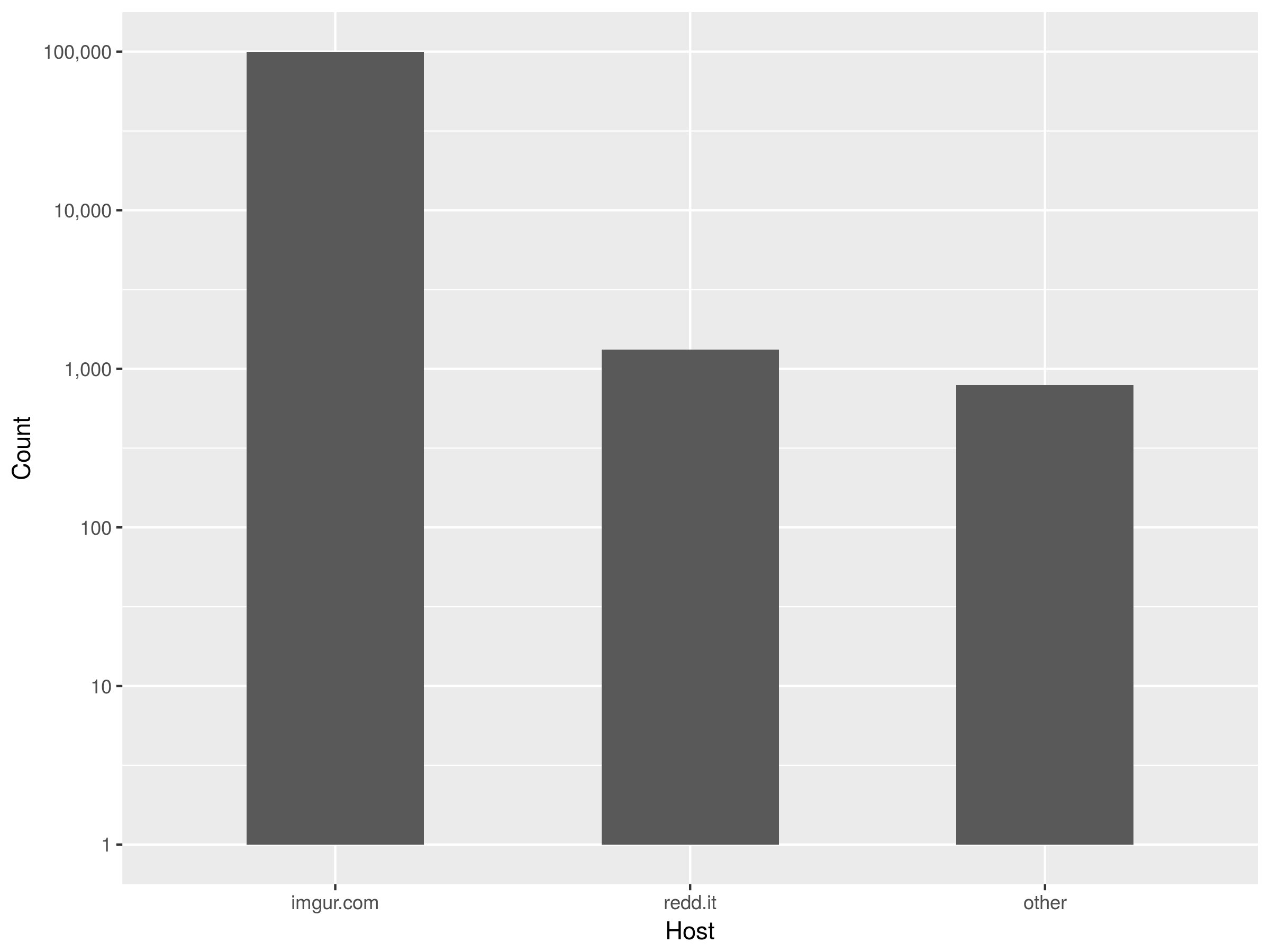}
\caption{Distribution of TLDs}
\label{fig:tld}
\end{figure}

We only considered posts and comments with a score above 20 to filter spam which was not already caught by community moderation and to ensure a minimal quality of manipulations. Figure \ref{fig:tld} shows the distribution of top-level domains. For the purpose of creating a large dataset, only supporting \url{imgur.com} would have been sufficient.
All images except one were in PNG or JPEG format. We cut the one image with WEBP format to reduce processing complexity. 
We left out all images which when crawled had a bytesize of less than 10~kB since those are mostly either removed images or thumbnails whose quality is too low for meaningful analysis.
%Images from an imgur gallery were also left out (albums are included)
%Comments which contained more than one linked URL were not included in the dataset as those might link sources for their derivates which diminishes dataset quality

\subsection{Structure}

The git repository mentioned above contains two primary sources of metadata -- \emph{originals.tsv} and \emph{photoshops.tsv} -- describing the original images and their resulting photoshops, respectively. The metadata for the original images contains the image URL, a unique id, the file size in bytes, a reference to the reddit post where the image was used, the username of the post, the community score as contained in the datadump, and image dimensions. It also contains a checksum to validate if the image was downloaded correctly. The metadata for the derived images has the same structure but additionally contains for every derivative the id of the original image. Instead of referencing the corresponding reddit post, the metadata for the photoshops references the relevant top-level comment where the image was posted.

Running the provided \emph{download.sh} script will use these metadata files in order to obtain the images from their respective sources and place them into the \emph{dataset} directory. The original images will be placed into the \emph{originals} subdirectory and renamed in accordance with their id while the photoshops will end up in a directory corresponding to the id of their original image beneath \emph{dataset/photoshops}.
Therefore, all derivations of the image \\ \emph{dataset/originals/(id).(filetype)} will be located in the directory \\ \emph{dataset/photoshops/(id)/}.

\subsection{Discussion} 
As the \emph{photoshopbattles} community is very active, new iterations of the dataset will grow in size. Very recently, the community has started to accept GIF manipulations as submissions. Including those in a new version of the dataset would be  interesting for the domain of video tampering detection which is not discussed here.
Another possibility for new versions of the dataset is manually filtering the comment chains for derivates of the created photoshops as the authors of \cite{moreira2018image} have done on a small subset of photoshops.

\section{Conclusion}
\label{sec:conclusion}

In this paper, we introduced the PS-Battles dataset. It contains \pscount{} original images and \derivatecount{} derivates of those images from the photoshopbattles subreddit.
The dataset is intended to provide a long-lasting benchmark for image tampering detection and derivate detection methods. Given the wide range of derivates regarding semantics, methods, and skill we expect the dataset to provide a significant challenge for tampering detection methods.

\section*{Acknowledgements}
This work was partly supported by the Chist-Era project IMOTION with contributions from the Swiss National Science Foundation (SNSF, contract no.\ 20CH21\_151571).

\balance 

\bibliographystyle{plain}
\bibliography{bibliography}

\end{document}